\documentclass[aps,prd,reprint,superscriptaddress,showpacs,showkeys]{revtex4-1}

\usepackage{graphicx}
\usepackage{epstopdf}
\usepackage{subfigure}

\usepackage{natbib}
\usepackage{mathrsfs}
\usepackage{amsmath}
\usepackage{enumerate}
\usepackage{enumitem}
\setenumerate{itemindent=2em,leftmargin=0pt,itemsep=0pt,partopsep=0pt}
\setitemize[1]{itemindent=2em,leftmargin=0pt,itemsep=2pt,partopsep=0pt,topsep=0pt}

\begin{document}


\title{ Improved limits on solar axions and bosonic dark matter from the CDEX-1B experiment using the profile likelihood ratio method }

\author{Y. Wang}
\affiliation{Key Laboratory of Particle and Radiation Imaging (Ministry of Education) and Department of Engineering Physics, Tsinghua University, Beijing 100084}
\affiliation{Department of Physics, Tsinghua University, Beijing 100084}

\author{Q. Yue}
\altaffiliation [Corresponding author: ]{yueq@mail.tsinghua.edu.cn}
\affiliation{Key Laboratory of Particle and Radiation Imaging (Ministry of Education) and Department of Engineering Physics, Tsinghua University, Beijing 100084}

\author{S.K. Liu}
\altaffiliation [Corresponding author: ]{liusk@scu.edu.cn}
\affiliation{College of Physics, Sichuan University, Chengdu 610064}

\author{K.J. Kang}
\affiliation{Key Laboratory of Particle and Radiation Imaging (Ministry of Education) and Department of Engineering Physics, Tsinghua University, Beijing 100084}

\author{Y.J. Li}
\affiliation{Key Laboratory of Particle and Radiation Imaging (Ministry of Education) and Department of Engineering Physics, Tsinghua University, Beijing 100084}

\author{H.P. An}
\affiliation{Key Laboratory of Particle and Radiation Imaging (Ministry of Education) and Department of Engineering Physics, Tsinghua University, Beijing 100084}
\affiliation{Department of Physics, Tsinghua University, Beijing 100084}

\author{J.P. Chang}
\affiliation{NUCTECH Company, Beijing 100084}

\author{J.H. Chen}
\affiliation{Institute of Physics, Academia Sinica, Taipei 11529}

\author{Y.H. Chen}
\affiliation{YaLong River Hydropower Development Company, Chengdu 610051}

\author{J.P. Cheng}
\affiliation{Key Laboratory of Particle and Radiation Imaging (Ministry of Education) and Department of Engineering Physics, Tsinghua University, Beijing 100084}

\author{W.H. Dai}
\affiliation{Key Laboratory of Particle and Radiation Imaging (Ministry of Education) and Department of Engineering Physics, Tsinghua University, Beijing 100084}

\author{Z. Deng}
\affiliation{Key Laboratory of Particle and Radiation Imaging (Ministry of Education) and Department of Engineering Physics, Tsinghua University, Beijing 100084}

\author{X.P. Geng}
\affiliation{Key Laboratory of Particle and Radiation Imaging (Ministry of Education) and Department of Engineering Physics, Tsinghua University, Beijing 100084}

\author{H. Gong}
\affiliation{Key Laboratory of Particle and Radiation Imaging (Ministry of Education) and Department of Engineering Physics, Tsinghua University, Beijing 100084}

\author{P. Gu}
\affiliation{College of Physics, Sichuan University, Chengdu 610064}

\author{X.Y. Guo}
\affiliation{YaLong River Hydropower Development Company, Chengdu 610051}

\author{H.T. He}
\affiliation{College of Physics, Sichuan University, Chengdu 610064}

\author{L. He}
\affiliation{NUCTECH Company, Beijing 100084}

\author{S.M. He}
\affiliation{YaLong River Hydropower Development Company, Chengdu 610051}

\author{J.W. Hu}
\affiliation{Key Laboratory of Particle and Radiation Imaging (Ministry of Education) and Department of Engineering Physics, Tsinghua University, Beijing 100084}

\author{H.X.~Huang}
\affiliation{Department of Nuclear Physics, China Institute of Atomic Energy, Beijing 102413}

\author{T.C. Huang}
\affiliation{Sino-French Institute of Nuclear and Technology, Sun Yat-sen University, Zhuhai, 519082}

\author{L.P. Jia}
\affiliation{Key Laboratory of Particle and Radiation Imaging (Ministry of Education) and Department of Engineering Physics, Tsinghua University, Beijing 100084}

\author{H.B. Li}
\affiliation{Institute of Physics, Academia Sinica, Taipei 11529}

\author{H. Li}
\affiliation{NUCTECH Company, Beijing 100084}

\author{M.X. Li}
\affiliation{College of Physics, Sichuan University, Chengdu 610064}

\author{J.M. Li}
\affiliation{Key Laboratory of Particle and Radiation Imaging (Ministry of Education) and Department of Engineering Physics, Tsinghua University, Beijing 100084}

\author{J. Li}
\affiliation{Key Laboratory of Particle and Radiation Imaging (Ministry of Education) and Department of Engineering Physics, Tsinghua University, Beijing 100084}

\author{X. Li}
\affiliation{Department of Nuclear Physics, China Institute of Atomic Energy, Beijing 102413}

\author{X.Q. Li}
\affiliation{School of Physics, Nankai University, Tianjin 300071}

\author{Y.L.~Li}
\affiliation{Key Laboratory of Particle and Radiation Imaging (Ministry of Education) and Department of Engineering Physics, Tsinghua University, Beijing 100084}

\author{B. Liao}
\affiliation{College of Nuclear Science and Technology, Beijing Normal University, Beijing 100875 }

\author{F.K. Lin}
\affiliation{Institute of Physics, Academia Sinica, Taipei 11529}

\author{S.T. Lin}
\affiliation{College of Physics, Sichuan University, Chengdu 610064}

\author{Y.D. Liu}
\affiliation{College of Nuclear Science and Technology, Beijing Normal University, Beijing 100875 }

\author{Y.Y. Liu}
\affiliation{College of Nuclear Science and Technology, Beijing Normal University, Beijing 100875 }

\author{Z.Z. Liu}
\affiliation{Key Laboratory of Particle and Radiation Imaging (Ministry of Education) and Department of Engineering Physics, Tsinghua University, Beijing 100084}

\author{H. Ma}
\affiliation{Key Laboratory of Particle and Radiation Imaging (Ministry of Education) and Department of Engineering Physics, Tsinghua University, Beijing 100084}

\author{Q.Y. Nie}
\affiliation{Key Laboratory of Particle and Radiation Imaging (Ministry of Education) and Department of Engineering Physics, Tsinghua University, Beijing 100084}

\author{J.H. Ning}
\affiliation{YaLong River Hydropower Development Company, Chengdu 610051}

\author{H. Pan}
\affiliation{NUCTECH Company, Beijing 100084}

\author{N.C. Qi}
\affiliation{YaLong River Hydropower Development Company, Chengdu 610051}

\author{C.K. Qiao}
\affiliation{College of Physics, Sichuan University, Chengdu 610064}

\author{J. Ren}
\affiliation{Department of Nuclear Physics, China Institute of Atomic Energy, Beijing 102413}

\author{X.C.~Ruan}
\affiliation{Department of Nuclear Physics, China Institute of Atomic Energy, Beijing 102413}

\author{V. Sharma}
\affiliation{Institute of Physics, Academia Sinica, Taipei 11529}
\affiliation{Department of Physics, Banaras Hindu University, Varanasi 221005}

\author{Z. She}
\affiliation{Key Laboratory of Particle and Radiation Imaging (Ministry of Education) and Department of Engineering Physics, Tsinghua University, Beijing 100084}

\author{L.~Singh}
\affiliation{Institute of Physics, Academia Sinica, Taipei 11529}
\affiliation{Department of Physics, Banaras Hindu University, Varanasi 221005}

\author{M.K. Singh}
\affiliation{Institute of Physics, Academia Sinica, Taipei 11529}
\affiliation{Department of Physics, Banaras Hindu University, Varanasi 221005}

\author{T.X. Sun}
\affiliation{College of Nuclear Science and Technology, Beijing Normal University, Beijing 100875 }

\author{C.J. Tang}
\affiliation{College of Physics, Sichuan University, Chengdu 610064}

\author{W.Y. Tang}
\affiliation{Key Laboratory of Particle and Radiation Imaging (Ministry of Education) and Department of Engineering Physics, Tsinghua University, Beijing 100084}

\author{Y. Tian}
\affiliation{Key Laboratory of Particle and Radiation Imaging (Ministry of Education) and Department of Engineering Physics, Tsinghua University, Beijing 100084}

\author{G.F. Wang}
\affiliation{College of Nuclear Science and Technology, Beijing Normal University, Beijing 100875 }

\author{L. Wang}
\affiliation{Department of Physics, Beijing Normal University, Beijing 100875}

\author{Q. Wang}
\affiliation{Key Laboratory of Particle and Radiation Imaging (Ministry of Education) and Department of Engineering Physics, Tsinghua University, Beijing 100084}

\author{Z. Wang}
\affiliation{College of Physics, Sichuan University, Chengdu 610064}

\author{H.T. Wong}
\affiliation{Institute of Physics, Academia Sinica, Taipei 11529}

\author{S.Y. Wu}
\affiliation{YaLong River Hydropower Development Company, Chengdu 610051}

\author{Y.C. Wu}
\affiliation{Key Laboratory of Particle and Radiation Imaging (Ministry of Education) and Department of Engineering Physics, Tsinghua University, Beijing 100084}

\author{H.Y. Xing}
\affiliation{College of Physics, Sichuan University, Chengdu 610064}

\author{Y. Xu}
\affiliation{School of Physics, Nankai University, Tianjin 300071}

\author{T. Xue}
\affiliation{Key Laboratory of Particle and Radiation Imaging (Ministry of Education) and Department of Engineering Physics, Tsinghua University, Beijing 100084}

\author{Y.L. Yan}
\affiliation{College of Physics, Sichuan University, Chengdu 610064}

\author{L.T. Yang}
\affiliation{Key Laboratory of Particle and Radiation Imaging (Ministry of Education) and Department of Engineering Physics, Tsinghua University, Beijing 100084}

\author{N. Yi}
\affiliation{Key Laboratory of Particle and Radiation Imaging (Ministry of Education) and Department of Engineering Physics, Tsinghua University, Beijing 100084}

\author{C.X. Yu}
\affiliation{School of Physics, Nankai University, Tianjin 300071}

\author{H.J. Yu}
\affiliation{NUCTECH Company, Beijing 100084}

\author{J.F. Yue}
\affiliation{YaLong River Hydropower Development Company, Chengdu 610051}

\author{X.H. Zeng}
\affiliation{YaLong River Hydropower Development Company, Chengdu 610051}

\author{M. Zeng}
\affiliation{Key Laboratory of Particle and Radiation Imaging (Ministry of Education) and Department of Engineering Physics, Tsinghua University, Beijing 100084}

\author{Z. Zeng}
\affiliation{Key Laboratory of Particle and Radiation Imaging (Ministry of Education) and Department of Engineering Physics, Tsinghua University, Beijing 100084}

\author{B.T. Zhang}
\affiliation{Key Laboratory of Particle and Radiation Imaging (Ministry of Education) and Department of Engineering Physics, Tsinghua University, Beijing 100084}

\author{F.S. Zhang}
\affiliation{College of Nuclear Science and Technology, Beijing Normal University, Beijing 100875 }

\author{L. Zhang}
\affiliation{College of Physics, Sichuan University, Chengdu 610064}

\author{Z.Y. Zhang}
\affiliation{Key Laboratory of Particle and Radiation Imaging (Ministry of Education) and Department of Engineering Physics, Tsinghua University, Beijing 100084}

\author{M.G. Zhao}
\affiliation{School of Physics, Nankai University, Tianjin 300071}

\author{J.F. Zhou}
\affiliation{YaLong River Hydropower Development Company, Chengdu 610051}

\author{Z.Y. Zhou}
\affiliation{Department of Nuclear Physics, China Institute of Atomic Energy, Beijing 102413}

\author{J.J. Zhu}
\affiliation{College of Physics, Sichuan University, Chengdu 610064}

\collaboration{CDEX Collaboration}
\noaffiliation



\date{\today}

\begin{abstract}
We present the improved constraints on couplings of solar axions and more generic bosonic dark matter particles using 737.1 kg-days of data from the CDEX-1B experiment. The CDEX-1B experiment, located at the China Jinping Underground Laboratory, primarily aims at the direct detection of weakly interacting massive particles using a p-type point-contact germanium detector. We adopt the profile likelihood ratio method for analysis of data in the presence of backgrounds. An energy threshold of 160 eV was achieved, much better than the 475 eV of CDEX-1A with an exposure of 335.6 kg-days. This significantly improves the sensitivity for the bosonic dark matter below 0.8 keV among germanium detectors. Limits are also placed on the coupling $g_{Ae} < 2.48 \times 10^{-11}$ from Compton, bremsstrahlung, atomic-recombination and de-excitation channels and $g^{eff}_{AN} \times g_{Ae} < 4.14 \times 10^{-17}$ from a $^{57}$Fe M1 transition at 90\% confidence level.
\end{abstract}


\maketitle

\section{Introduction}
\par For the charge-parity (CP) problem of strong interactions, the Peccei-Quinn (PQ) mechanism~\cite{1977_PQ} is still the most compelling solution in which a new kind of U(1) symmetry would be spontaneously broken at large energy scale $f_A$. After this original solution to the CP conservation in QCD, a new Nambu-Glodstone boson called axion is proposed later by Weinberg~\cite{weinberg_axion} and Wilczek~\cite{wilczek_axion} through the PQ symmetry. Axions are pseudoscalar particles with properties closely related to those of neutral pions and their mass $m_A$ is fixed by the scale $f_A$ of the PQ symmetry breaking, $m_A$ $\approx$ 6 eV ( $10^6$ GeV/$f_A$ ). The range of  scale $f_A$ can not be restricted by theory but the order of the electroweak scale has been excluded by experiments. At a higher symmetry-breaking energy scale, `invisible' axion models such as hadronic model KSVZ (Kim-Shifman-Vainstein-Zakharov)~\cite{1979_Kim, 1980_Shifman} and non-hadronic model DFSZ (Dine-Fischler-Srednicki-Zhitnitskii)~\cite{1981_Dine, 1980_Zhitniskiy} are still allowed. Another interest in this paper is more general bosonic dark matter (DM) like axion-like particles (ALPs) and vector bosonic DM, which also have couplings to electrons.
\par Several experiments have reported the corresponding results \cite{cbrd_solar,axion_cdms,axion_cogent,cbrd_red,axion_edelweiss2,darkphoton_anhp,axion_majorana,axion_lux,axion_pandax2,axion_xenon_2017,c1a_axion,bosonic_xenon,axion_edelweiss3,texono,xenon_1t_axion,cos100} using the mechanism arising from the couplings to electrons:
\begin{equation}
A(B) + e + Z \rightarrow e + Z,
\label{eqae}
\end{equation}
where $A$ and $B$ represent axion and bosonic DM respectively. This effect is similar to photoelectric effect just replacing photon with axion or bosonic DM.
\par The China Dark Matter Experiment (CDEX) is primarily designed to carry out direct detection of low mass weakly interacting massive particles (WIMPs) with p-type point contact germanium detectors (PPCGe) at China Jinping Underground Laboratory (CJPL) \cite{CDEX_introduction,cjpl,cdex12014,cdex102018,cdex0vbb2017,cdex_migdal,ylt_anu}. With a vertical rock overburden of 2.4 km, CJPL provides a measured muon flux of 61.7$\pm$11.7 y$^{-1}$ m$^{-2}$ \cite{wuyc2013}. Besides the WIMPs constraints \cite{cdex1,cdex12016}, the axion searches results from the CDEX-1A experiment based on the 335.6 kg-days of data has been reported before \cite{c1a_axion}. Using a PPCGe with fiducial mass of 915 g, a physics threshold of 475 eV \cite{cdex12016} was achieved for CDEX-1A. Focused on the lower energy threshold, a new 1 kg-scale PPCGe detector has been designed and named `CDEX-1B' based on the experience from our previous prototype detector used in CDEX-1A.

In this paper, We report the solar axion, ALPs and vector bosonic DM searches results from the CDEX-1B experiment based on the 737.1 kg-days of data, which is the same data set in the analysis of WIMP search \cite{cdex1b2018}, annual modulation \cite{ylt_anu} and Midgal effects \cite{cdex_migdal}. Also we describe the statistical model with profile likelihood ratio method applied to this data.

\section{AXION SEARCHES WITH CDEX-1B}
\subsection{CDEX-1B setup and overview}
The CDEX-1B experiment adopts one 939 g single-element PPCGe crystal with dead layer of 0.88 $\pm$ 0.12 mm \cite{majinglu2017}. Outside of the PPCGe detector is the passive shielding system and the detailed information is described in Ref. \cite{cdex1b2018}. A well-shaped cylindrical NaI(Tl) crystal surrounding the PPCGe detector is used as the anti-Compton detector. The coincidence events both in germanium and NaI(Tl) crystals denoted as AC$^+$ are discarded to depress the $\gamma$ background.
 \par The schematic diagram of electronics and data acquisition (DAQ) system is shown in Ref. \cite{cdex1b2018}. Four identical energy-related signals were out of the p$^+$ point-contact electrode after a pulsed-reset feedback preamplifier. Two of them were distributed into the shaping amplifiers at 6 $\mu$s (SA$_{6{\mu}s}$) and 12 $\mu$s (SA$_{12{\mu}s}$) shaping time for low energy region (0-12 keV). The output of SA$_{6{\mu}s}$ provided the system trigger of the DAQ. The other two outputs were fed into timing amplifiers (TA) which provide the accurate time information. One with high gain (TA$_1$) is limited to medium energy region (0-20 keV), and the other one with low gain (TA$_2$) for high energy can reach 1.3 MeV. The energy resolution of TA$_1$ output is similar to the SA$_{6{\mu}s,12{\mu}s}$. As a result, the spectrum below 12 keV is from SA$_{6{\mu}s}$ and above 12 keV is from TA$_1$ in our analysis. The energy resolution ($\sigma$) from SA$_{6{\mu}s}$ at 1.3 keV is about 44 eV.
 \subsection{Particle sources}
 \subsubsection{Solar Axions}
 The sun is a potential source of axions and in this article we concentrate on two different mechanisms.

 \par The first important source is the 14.4 keV monochromatic axions from the M1 transition of the $^{57}$Fe in the sun, i.e. $^{57}$Fe$^{*}\rightarrow^{57}$Fe+A, due to the stability and the large abundance of $^{57}$Fe in the sun.

 \par The Lagrangian coupling axions to nucleons is \cite{axion_edelweiss2}:
 \begin{equation}
 \mathcal{L} = i{\bar{\psi}}_N{\gamma}_5(g_{AN}^0 + g_{AN}^3{\tau}_3){\psi}_N{\phi}_A,
 \end{equation}
where ${\psi}_N$ is the nucleon isospin doublet, ${\phi}_A$ is the axion field, and ${\tau}_3$ is Pauli matrix. $g_{AN}^0$ and $g_{AN}^3$ are the model-dependent isoscalar and isovector axion-nucleon coupling constants \cite{1985_kaplan,1985_srednicki}. Introducing $g_{AN}^{\rm eff} \equiv (-1.19g_{AN}^0+g_{AN}^3)$ as the effective nuclear coupling adapted to the case of $^{57}$Fe, the corresponding axion flux is given by \cite{axion_edelweiss2,cast_fe57}:
 \begin{equation}
 {\Phi}_{14.4}=\left(\frac{{\kappa}_A}{{\kappa}_{\gamma}}\right)^3 \times 4.56 \times 10^{23}(g_{AN}^{\rm eff})^2 \ {\rm cm}^{-2}{\rm s}^{-1},
 \end{equation}
 where ${\kappa}_A$ and ${\kappa}_{\gamma}$ are the momenta of the outgoing axion and photon respectively. Given the axion-nucleon couplings $g_{AN}^0$ and $g_{AN}^3$ for specific models such as DFSZ and KSVZ, the axion flux can be evaluated.

 \par Another important sources are from the Compton-like scattering (C), axion-bremsstrahlung (B), atomic-recombination (R) and atomic-deexcitation (D) processes. Their corresponding effective Lagrangian is given by \cite{axion_edelweiss2}:
 \begin{equation}
 \mathcal{L} = ig_{Ae}{\bar{\psi}}_e{\gamma}_5{\psi}_e{\phi}_A,
 \end{equation}
 where $g_{Ae}$ is the dimensionless axion-electron coupling constant. Its flux depends on the $g_{Ae}^2$:
 \begin{equation}
 \begin{aligned}
 &\frac{d{\Phi}_{CB}}{dE_A} = \frac{d{\Phi}_{C}}{dE_A} + \frac{d{\Phi}_{B}}{dE_A} \\
  &= g_{Ae}^2 \times 1.33 \times 10^{33}E_{A}^{2.987}e^{-0.776E_A} \\
  &+  g_{Ae}^2 \times 2.63 \times 10^{35}E_{A}e^{-0.77E_A} \frac{1}{1+0.667E_A^{1.278}},
 \end{aligned}
 \end{equation}
 where the units of fluxes are cm$^{-2}$s$^{-1}$keV$^{-1}$ and axion energy $E_A$ is in unit of keV. For the atomic-recombination and atomic-deexcitation process, the tabulated spectrum in Ref. \cite{spec_rd} is used. As discussed in Ref. \cite{spec_rd}, the flux is valid for relativistic axion; hence, we consider only the axion mass below 1 keV.

\par The axion-electron coupling is depended on models. In the DFSZ model, the coupling is proportional to $\rm cos^{2}{\beta}$, where $\rm tan{\beta}$ is the ratio of the two Higgs vacuum expectation values. In the KSVZ model, it depends on $E/N$, the ratio of electromagnetic to color anomalies. $E/N=0$ and $\rm cos^2{\beta}=1$ are used in this analysis \cite{axion_edelweiss2}.

\subsubsection{Bosonic Dark Matter}
The main cosmological interest in bosonic particles such as ALPs and vector bosonic DM arises from their possible role as the dominant component of dark matter, the nature of which is still unknown. The absorption via ionization or excitation of an electron in target atom makes bosonic DM experimentally interesting and PPCGe detectors have advantages to study bosonic DM due to their excellent energy resolution, sub$-$keV threshold and low radioactivity background.
\par Assuming that these bosonic particles constitute all of the galactic dark matter, we get the total average flux of dark matter axions on Earth:
\begin{equation}
\begin{aligned}
{\Phi}_{\rm DM} &= {\rho}_{\rm DM} \cdot v_A/m_A \\
            &= 9.0 \times 10^{15} \times \beta \cdot \left(\frac{\rm keV}{m_A}\right) {\rm cm}^{-2}{\rm s}^{-1},
\end{aligned}
\end{equation}
where ${\rho}_{DM}\sim 0.3$ GeV/{cm$^3$} is the dark matter halo density \cite{2012_Green}, $m_A$ is the axion mass, $v_A$ is the mean axion velocity distribution with respect to the Earth and $\beta$ is the ratio of the axion velocity to the speed of light for cold dark matter. This flux is independent of any axion coupling.

\subsection{Particle interactions in CDEX-1B}
The axion detection channel studied in this paper is the axio-electric effect illustrated in Eq. (1). The axio-electric cross-section as described in Ref. \cite{coure_fe57,2010_derevianko,2008_pospelov} is given by:
\begin{equation}
{\sigma}_{Ae}(m_A)={\sigma}_{pe}(m_A)\frac{g_{Ae}^2}{\beta} \frac{3{m_A}^2}{16\pi\alpha{m_e^2}} \left(1-\frac{{\beta}^{\frac{2}{3}}}{3} \right),
\end{equation}
where $\sigma_{pe}(m_A)$ is the photo-electric cross-section for germanium in the unit of barns/atom, $m_A$ is the mass of axion, $\alpha$ is the fine structure constant, $m_e$ is the electron mass and $\beta$ is the ratio of the axion velocity to the speed of light. The expected axion event rates of CBRD process and $^{57}$Fe under the consideration of energy resolution are displayed in the Fig. 1. 

\par In the situation of ALPs in cold dark matter model ($\beta\approx10^{-3}$), the coupling to electrons is the same as in the case of solar axions. For the vector bosonic DM, the absorption cross section $\sigma_{\rm abs}$ can be written as:
\begin{equation}
{\sigma}_{\rm abs}(m_v)={\sigma}_{pe}(m_v)\frac{{\alpha}'}{\alpha},
\end{equation}
where $m_{v}$ is the mass of the vector bosonic DM, $\alpha$ and ${\alpha}'$ are the fine structure constant and its vector boson equivalent, respectively.

Using the parameter mentioned above, the interaction rate in the direct detection experiment can be written as:
\begin{equation}
 R=1.2\times10^{43}A^{-1}g_{Ae}^{2}m_{A}\sigma_{pe}(m_A)
\end{equation}
for ALPs and
\begin{equation}
 R=4\times10^{47}A^{-1}\frac{{\alpha}'}{\alpha}m_v^{-1}\sigma_{pe}(m_v)
\end{equation}
for vector bosonic DM, where $A$ is mass number of germanium. The expected rates of these two kinds of particles are shown in Fig. 2.

\begin{figure}[htb]
 \includegraphics[width=1.0\linewidth]{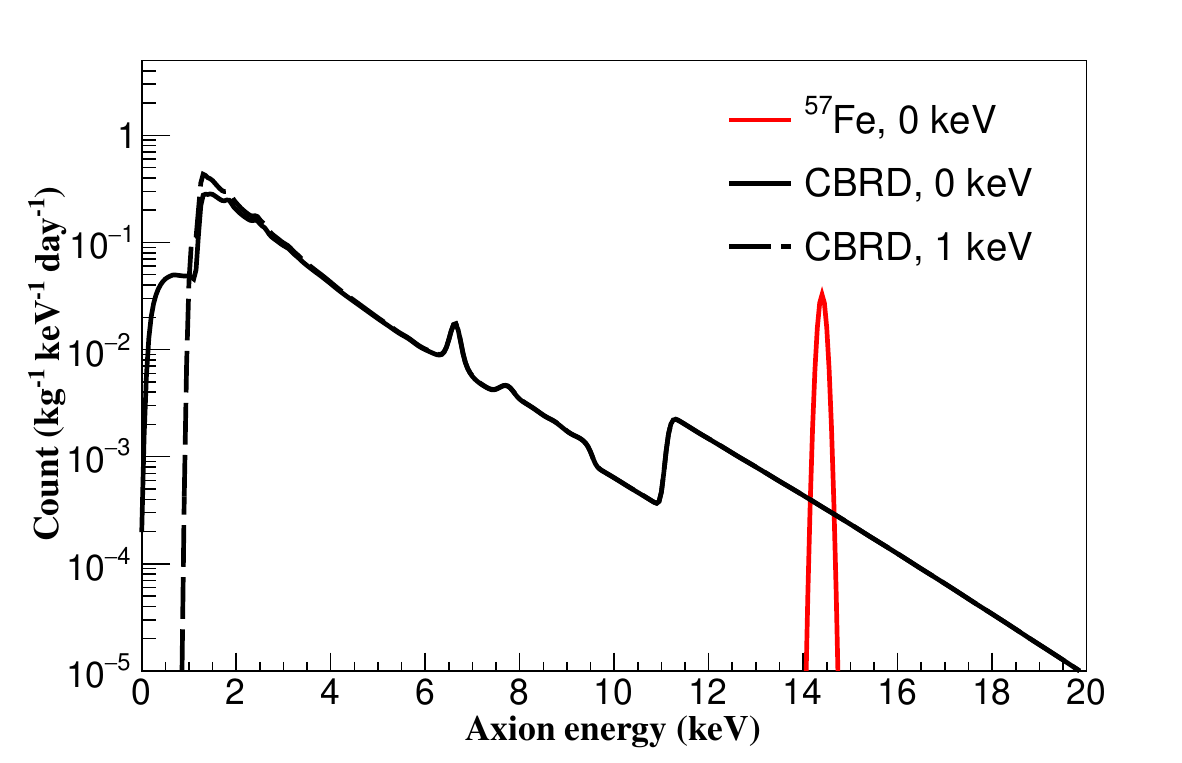}
  \caption{\label{fig:solar_rate} The expected axion event rates of CBRD process at the mass of 0 keV and 1 keV, and $^{57}$Fe 14.4 keV axion at the mass of 0 keV. Here the axion couplings are $g_{Ae}=2\times10^{-11}$ and $g_{AN}^{eff}\times{g_{Ae}}=2\times10^{-17}$.}
 \end{figure}

\begin{figure}[htb]
 \includegraphics[width=1.0\linewidth]{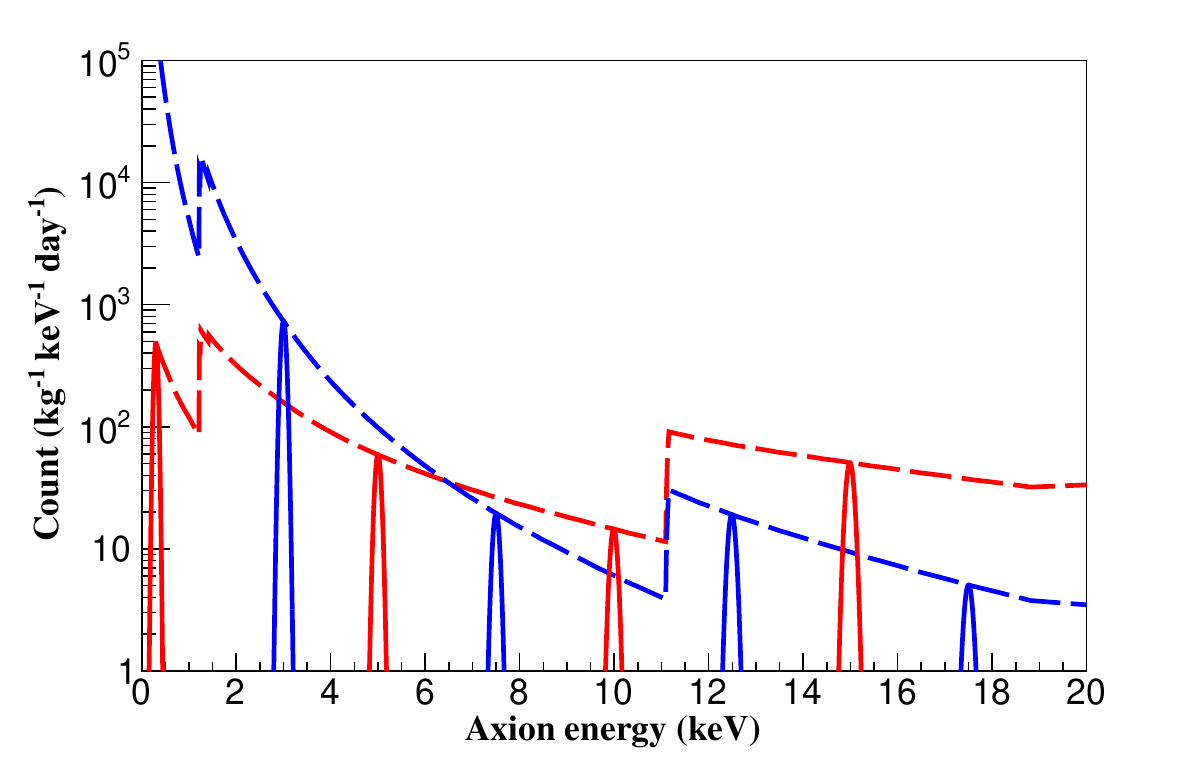}
  \caption{\label{fig:dm_rate} The expected event rate of ALPs (red solid line) and vector bosonic DM (blue solid line) at different masses. The red dashed line is the maximum event rate of ALPs Gaussian distributions versus their masses, while the blue dashed line is corresponding to vector bosonic DM. The couplings used  here are $g_{Ae}=2\times10^{-11}$ and ${\alpha}'/\alpha=5\times10^{-25}$. The widths of these peaks are determined by the energy resolution.}
 \end{figure}

\section{DATA ANALYSIS}
\subsection{Data Selection}
As discussed in earlier analysis \cite{cdex1b2018}, the background spectrum is derived by the following steps:
\begin{enumerate}
\setlength{\itemsep}{0.5ex}
  \item Stability check, removing the time periods of calibration or other testing experiments.
  \item Anti-Compton (AC) veto, discarding the events in coincidence with the AC detector and retaining the anti-coincidence events.
  \item Basic cuts, removing the electronic noise through getting rid of the abnormal pulses and spurious signals.
  \item Bulk and surface event selection, rejecting the surface events by pulse shape analysis using their characteristic slower rise time.
 \end{enumerate}
 \par
\begin{figure}[htb]
 \includegraphics[width=1.0\linewidth]{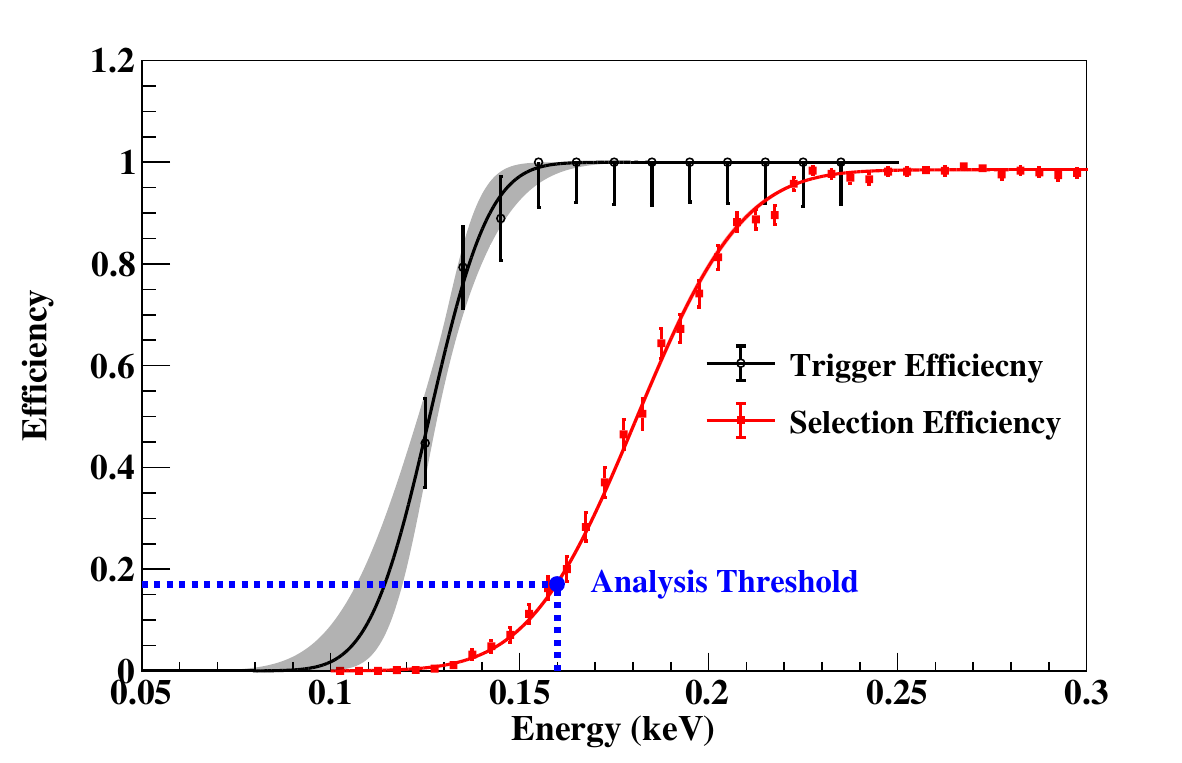}
  \caption{\label{fig:efficiency}The trigger efficiency and selection efficiency derived from source events are depicted respectively and fitted with error functions. The shadow
parts represent the 1$\sigma$ error bands.}
 \end{figure}

  Depicted in Fig.~\ref{fig:efficiency} are the trigger efficiency as well as the selection efficiency with energy including those from the selection of physics vs electronic noise events, AC vetos and DAQ dead time. The trigger efficiency is derived from the calibration sources in coincidence with AC detector \cite{cdex1b2018}. The selection efficiencies are derived by events due to random triggers, the AC tagged events from calibration sources and in situ background. An improved Ratio Method, which is based on the bulk/surface rise time distribution probability density functions (PDFs), is developed to reject the surface events \cite{yanglt_bs}. This method has been proved correctly above 160 eV. So in this analysis, 160 eV is selected as the physics analysis threshold, at which the combined efficiencies ($\varepsilon_{\rm eff}$) including trigger and selection is 17\%.

\subsection{Background and Understanding}
\begin{figure}[htb]
   \centering
   \begin{subfigure}[]
   \centering
    \label{fig:bkg}
   \includegraphics[width=1.0\linewidth]{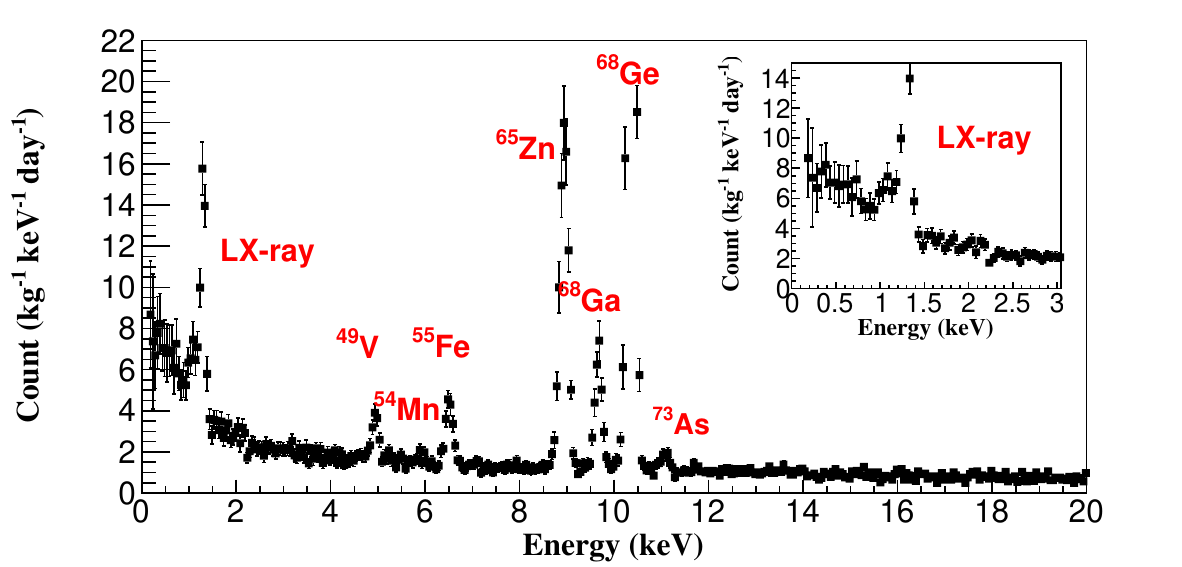}
   \end{subfigure}
   \begin{subfigure}[]
   \centering
   \label{fig:sr_spec}
   \includegraphics[width=1.0\linewidth]{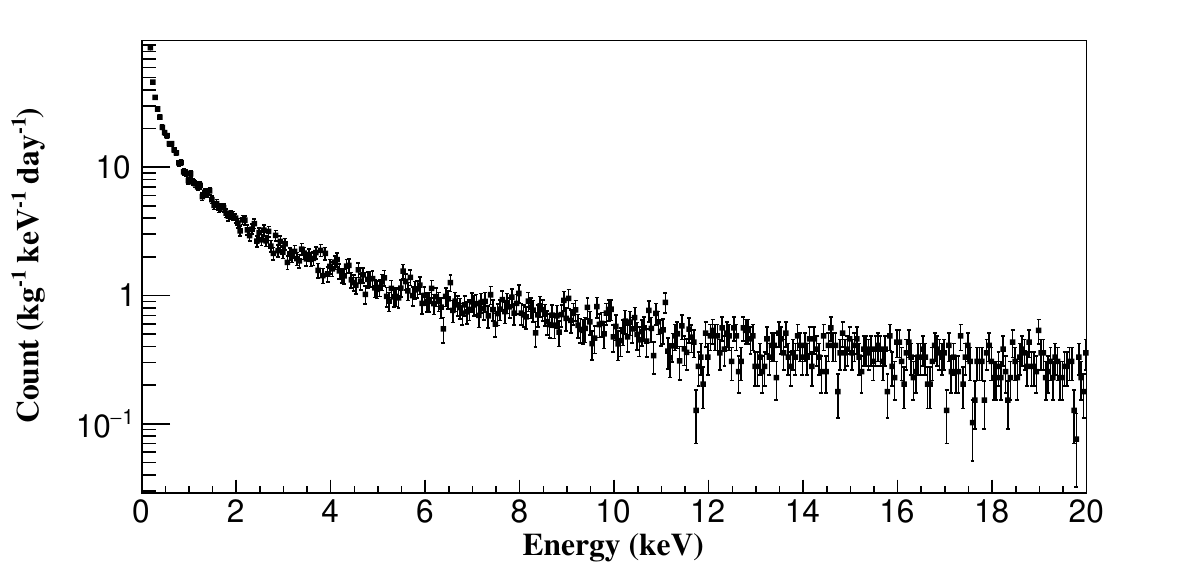}
   \end{subfigure}
   \caption{\label{fig:corrected_spectrum} (a) The corrected bulk spectrum from 160 eV to 20 keV. (b) The corrected surface spectrum from 160 eV up to 20 keV.}
 \end{figure}

With an exposure of 737.1 kg-days, the bulk spectrum from 160 eV up to 20 keV after data selection and efficiency correction is displayed in Fig.~\ref{fig:corrected_spectrum}(a). The background consists of several K-shell X-rays and their corresponding L-shell X-rays from the cosmogenic isotopes and a continuous background with a smooth, slightly increasing profile as the energy decreases \cite{cdex1b2018}. Considering the low muon flux mentioned above, the contribution from muons can be neglected. The continuous background below 20 keV is expected to probably originate from the the $^{238}$U, $^{232}$Th and $^{40}$K in the materials in the vicinity of the PPCGe detector, radon gas penetrating through shielding and cosmogenic $^3$H in the crystal. A detailed modeling of the continuous background is beyond this work and will be studied in our future work.
\par However, axion analysis is not sensitive to the accurate background assumption because the signatures of axion are significantly different from the continuous background. As can be seen from Fig.~\ref{fig:dm_rate}, the signal signatures of $^{57}$Fe and bosonic DM are monochromatic and of Gaussian distribution with widths determined by the energy resolution. As to the continuous CBRD solar axion, a saw-tooth-like profile arises between 0.9 keV and 1.6 keV considering the axion mass below 1 keV. So in the following fitting procedure, the background model can be described by a continuous background plus the peaks from K/L-shell X-rays. Benefiting from the low threshold and excellent energy resolution of CDEX-1B, the L-shell X-ray peaks at low energy region can be clearly distinguished. Therefore, in the background model, the amplitude of the K-shell X-ray peaks and the corresponding L-shell X-ray peaks are limited by each other using the K/L-shell X-ray ratios mentioned in Ref. \cite{1963_kl,1976_kl}. In the ultra low energy region around the threshold, M-shell X-rays are also taken into consideration in the background model.

The corrected surface spectrum derived from Ratio Method is depicted in Fig.~\ref{fig:corrected_spectrum}(b). Note that, as will be clear in next section, the likelihood analysis makes use of
both the bulk and surface data.

\subsection{Profile likelihood Analysis}
A profile likelihood analysis, as described in Ref. \cite{profilelikelihood}, is adopted to derive the constraints and the test statistics is:
 \begin{equation}
 q_{\mu}=\left\{
             \begin{array}{lcl}
             -2ln\left( \frac{L\left(\mu,\hat{\hat{\theta}} \right)}{L\left(\hat{\mu},\hat{\theta}  \right)}\right) \:\:\: \mu\geq\hat{\mu} & \\
             0 \:\:\:\:\:\:\:\:\:\:\:\:\:\: \:\:\:\:\:\:\:\:\:\:\:\:\:\:\:\:\:\:    \mu<\hat{\mu}, &
             \end{array}
           \right.
\end{equation}
where $L$ is the likelihood function. Quantity $\mu$ is a parameter corresponding to the strength of signals and $\theta$ denotes all of the nuisance parameters. The quantity $\hat{\hat{\theta}}$ denotes the value of $\theta$ that maximizes $L$ for the specified $\mu$, while the denominator is the maximized likelihood function, i.e. $\mu$ and $\theta$ are their maximum-likelihood estimators. To obtain the 90\% C.L. bounds on the signal strengths $\mu$,
the asymptotic formulas are used to calculate the probability distribution functions (PDFs), i.e., 

\begin{equation}
\begin{aligned}
 f(q_{\mu}|\mu')&=\Phi(\frac{\mu'-\mu}{\sigma})\delta(q_{\mu})\\
                &+\frac{1}{2}\frac{1}{\sqrt{2\pi}}\frac{1}{\sqrt{q_{\mu}}}\rm exp[-\frac{1}{2}(\sqrt{q_{\mu}}-\frac{\mu-\mu'}{\sigma})^2],
\end{aligned}
\end{equation}
where $f(q_{\mu}|\mu')$ is the PDF of the test statistic $q_{\mu}$ under the signal strength hypothesis $\mu'$, while $\sigma$ is the corresponding standard deviation \cite{profilelikelihood}. Since downward fluctuations of background might lead to much stringent exclusion results, we used the CL$_s$ method \cite{clsmethod} to get rid of this effect. The 90\% up limits $\mu_{\rm up}$ are defined as:
\begin{equation}
 \frac{1-F(q_{\mu}|{\mu})_{\rm up}}{1-F(q_{\mu}|0)_{\rm up}}=10\%,
\end{equation}
where $F$ is the cumulative distribution function of the test statistic.

\subsubsection{Likelihood Function}

\begin{figure}[htb]
   \centering
   \begin{subfigure}[]
   \centering
   \label{fig:2d_spec}
   \includegraphics[width=1.0\linewidth]{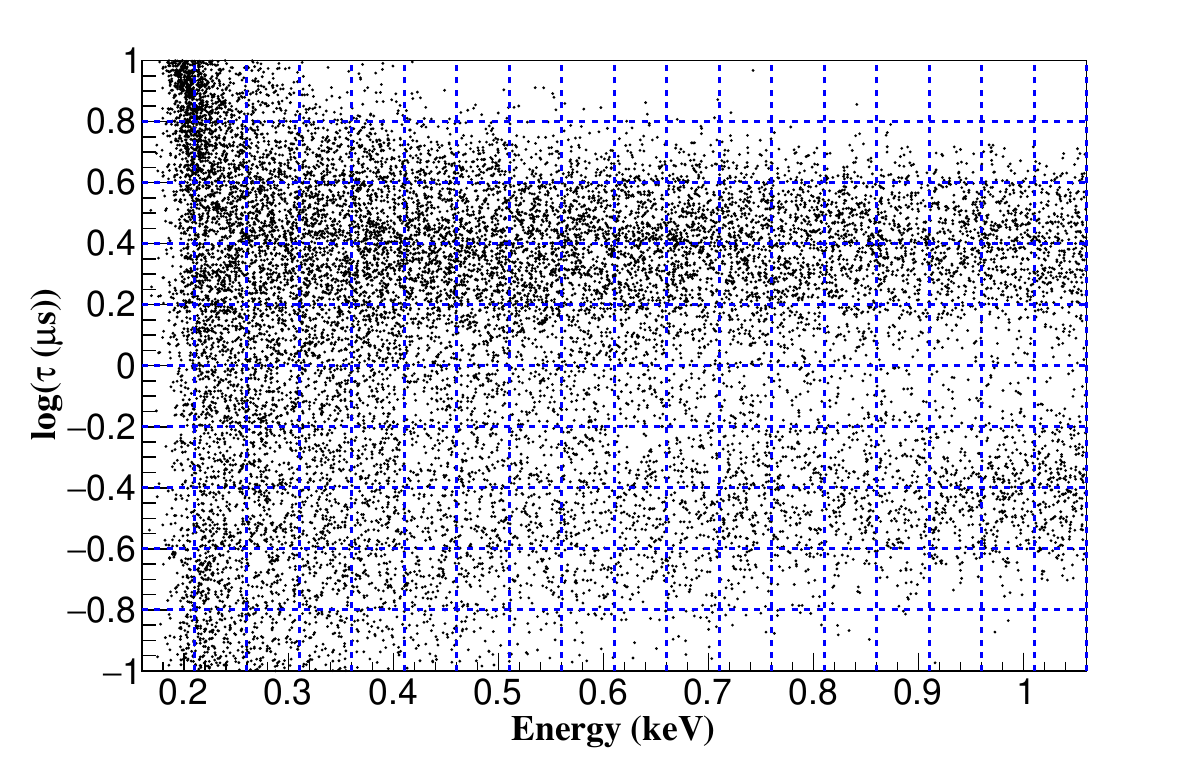}
   \end{subfigure}
   \begin{subfigure}[]   
   \centering
   \label{brsr}
   \includegraphics[width=1.0\linewidth]{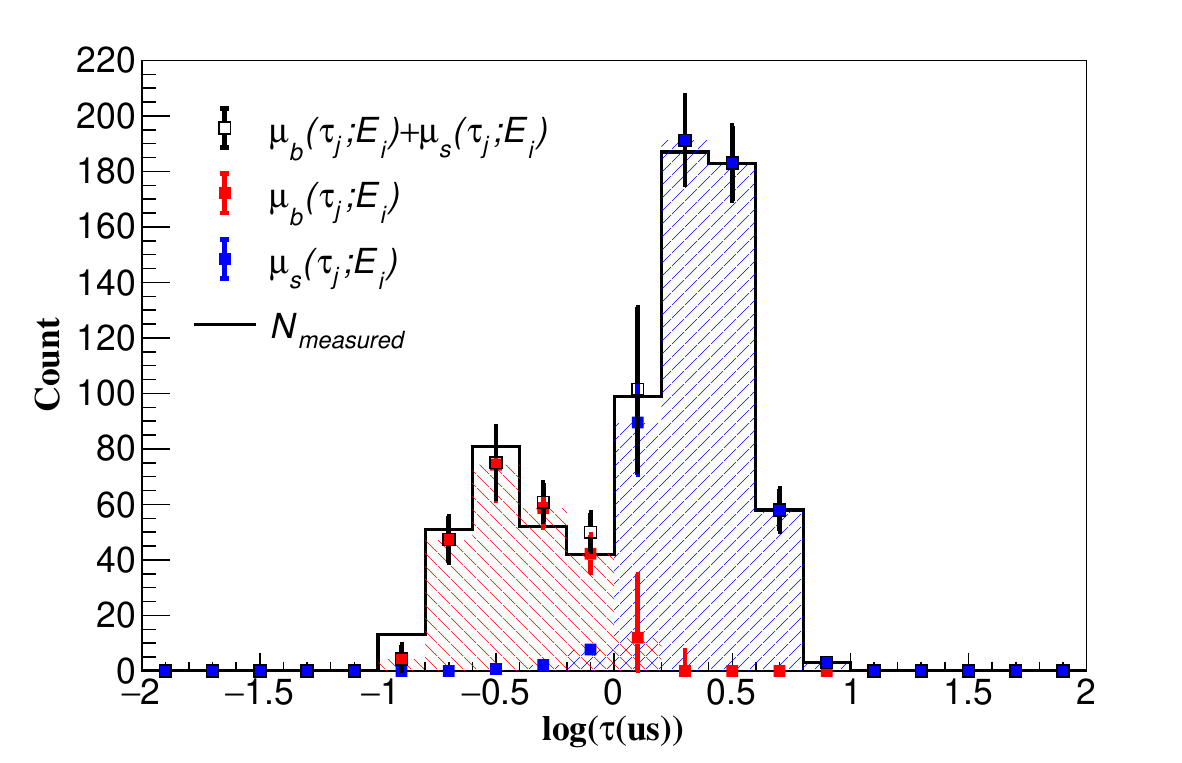}
   \end{subfigure}
   \caption{\label{fig:riset_distribution} (a) The event distribution in the rise time vs energy and the blue dashed grid displayed here shows the method of binning.  (b) The distribution of rise time in energy bin 0.66 to 0.71 keV. The $\mu_{b}$ and $\mu_{s}$ are the normalized PDFs of bulk and surface events in this energy bin derived from the output of Ratio Method and properly scaled to be compared with the measured numbers. }
 \end{figure}

The specific full likelihood function  $ \mathcal{L} $ we used in this analysis is written as a product of three terms:
\begin{equation}
\begin{aligned}
 \mathcal{L} = \mathcal{L}_1 (\nu_{A}, \nu_{b}, \nu_{s},  g_{b}, g_{s}, \varepsilon_{\text{eff}}; m_{A} ) \times \mathcal{L}_2( \varepsilon_{\text{eff}}(E))\\
 \times \mathcal{L}_3(t_{b},t_{s}) ,
 \end{aligned}
\end{equation}
the parameter of interest becomes the number of fitted axion event number denoted $\nu_A$ which is related to the axion-electron coupling strength $g_{Ae}$, whereas $\nu_{b}, \nu_{s}, g_{b}, g_{s}, \varepsilon_{\text{eff}}$ are considered as the main nuisance parameters.

\begin{equation}
 \begin{aligned}
 \mathcal{L}_{1} = \prod_{j=1}^{N_{\tau}}\prod_{i=1}^{N_{E}}{\rm Poisson}[n_{ij}|(g_{b}(\tau_j;E_{i};t_{b})\cdot \varepsilon_{\rm eff}(E_{i})\cdot N_{i,{\rm bulk}}\\
  +g_{s}(\tau_j;E_i;t_{s})\cdot \varepsilon_{\rm eff}(E_{i})\cdot N_{i,{\rm surf}})]
 \end{aligned}
\end{equation}
describes the measurement of the detector.  Here we projected all the data into the Energy versus rise-time 2-dimension grids, as depicted in Fig.~\ref{fig:riset_distribution}(a). The $n_{ij}$ is the measured event number both in the energy spectrum bin $E_{i}$ and the rise time spectrum bin $\tau_{j}$. $g_{b}(\tau;E_{i};t_{b})$ and $g_{s}(\tau;E_{i};t_{s})$ are the distributions of rise-time at the condition of a certain energy bin $i$ from bulk event and surface event respectively, i.e., $ g_{k}(\tau_j; E_i)= \mu_{k}(\tau_j; E_i)+t_{k}\cdot\sigma_{k, ij}$, $k=$ bulk or surface.
Normalized PDFs $\mu_{k}(\tau_j; E_i)$ are the best fit values derived from Ratio Method in the rise-time distribution shown in Fig.~\ref{fig:riset_distribution}(b), as well as their corresponding errors $\sigma_{k, ij}$ including statistical and systematic uncertainties which have already been derived in Ref. \cite{yanglt_bs,cdex1b2018}.
$\varepsilon_{\rm eff}$ described by $\vec{e}$ refers to the combined efficiencies mentioned in the Sec. III. A: 
\begin{equation}
 \begin{aligned}
   \varepsilon_{\rm eff}&=\left\{\frac{1}{2}\times[1+{\rm Erf}(\frac{E-e_1}{\sqrt{2}e_2})]\right\}_{\rm trigger}\\
                    &\times\left\{\frac{e_3}{2}\times[1+{\rm Erf}(\frac{E-e_4}{\sqrt{2}e_5})]\right\}_{\rm selection}
  \end{aligned}
\end{equation}

$N_{i,{\rm bulk}}$ and $N_{i,{\rm surf}}$ are the expected numbers of bulk events and surface events at the certain energy bin $i$, respectively, which are determined by the fitting results:
\begin{equation}
 \begin{aligned}
 N_{i,{\rm bulk}}=\nu_{b} \cdot f_{b}(E_i) +\nu_{A} \cdot f_{A}(E_i)\\,
 N_{i,{\rm surf}}= \nu_{s} \cdot f_{s}(E_i).
   \end{aligned}
\end{equation}
$f_{b}, f_A$ and $f_s$ represent the PDFs of the background, the axion signal and the surface events, respectively. Each of them is normalized to unity over the energy range of the fit. 
$f_A$ describes the axion events as shown in Fig.~\ref{fig:solar_rate} and Fig.~\ref{fig:dm_rate}.
The background $f_{b}$ consists of K-shell X-ray peaks from the cosmogenic nuclides and their corresponding L-shell X-rays and a continuous component with a smooth, slightly increasing profile as the energy decreases. 
The surface event $f_s$ is derived from fitting the surface spectrum with a smooth curve. The systematic uncertainties of the PDF selection of  $f_{s}$ is negligible by comparing bin by bin PDFs from the  $f_{s}$ spectrum. The number of surface events derived from the Ratio Method is used as $\nu_s$ and fixed in the likelihood fit. The results are consistent with the situation in which $\nu_s$ is free, but more conservative below 400 eV in the bosonic DM fit. While $\nu_b$ and $\nu_A$ fitted as free parameters are the numbers of background events and axion events, respectively.

\subsubsection{Constraints and Systematic uncertainties}

$ \mathcal{L}_{2}$ is a constraint term which encodes prior constraints on the combined efficiencies $\varepsilon_{\rm eff}$,
\begin{equation}
\begin{aligned}
\mathcal{L}_2 &={\rm exp}[ {-\frac{1}{2}\sum_{i,j=1}^{2}(e_i-\mu_{ei})\emph{\textbf{V}}_{ij}^{-1}(e_j-\mu_{ej})}]\\
              &\times{\rm exp}[ {-\frac{1}{2}\sum_{i,j=3}^{5}(e_i-\mu_{ei})\emph{\textbf{V}}_{ij}^{-1}(e_j-\mu_{ej})}].
\end{aligned}
\end{equation}
The five parameters $\vec{e}$ used in two error functions to describe the trigger efficiency and selection efficiency included in $\varepsilon_{\rm eff}$ are constrained by $ \mathcal{L}_{2}$, with 2D and 3D Gaussians respectively. Both centers of the Gaussians are derived by the best-fit values of parameters denoted $\vec{\mu_{e}}$ depicted in Fig.~\ref{fig:efficiency} , and their shapes are determined by the covariance matrix  $\emph{\textbf{V}}$ between the best-fit values.

According to the evaluation in the previous work \cite{cdex1b2018}, one of the dominated uncertainties at the energy range below 1 keV, including statistical and systematic errors, originate from the bulk surface event selection, i.e., the nuisance parameters $g_k(\tau_j;E_{i})$, in likelihood function $\mathcal{L}_1$.
In order to take this uncertainties into consideration,  $\mathcal{L}_{3}$ term is introduced, 
\begin{equation}
 \mathcal{L}_3 = e^{-t_b^2/2} \times e^{-t_s^2/2},
\end{equation}
which has been parametrized with two parameters $t_b$, $t_s$. 
The likelihood function is defined to be a product of two normally Gaussian distributions, corresponding to where $t=\pm 1$ corresponds to a $\pm 1\sigma$ deviation in $g_k(\tau_j;E_{i})$.

The uncertainties of the background assumption $f_b$ are evaluated by using different continuous component in the background assumption between different combinations of exponential, polynominal and flat functions for the fit below 12 keV. For the energy range around 14.4 keV, background assumptions are varied between polynomial, flat and exponential function. The variation of background models causes the change of constraints less than 8\% for CBRD axion, less than 16\% for bosonic DM, and less than 8\% for $^{57}$Fe solar axion. As for the uncertainties of resolution, varying the energy resolution by $\pm10\%$, the changes of results are less than 17\% for $^{57}$Fe solar axion, less than 13\% for bosonic DM and negligible for CBRD axion.

\section{AXION SENSITIVITY ANALYSIS AND RESULTS}
\subsection{14.4 keV Solar Axion}
\begin{figure}[htb]
 \includegraphics[width=1.0\linewidth]{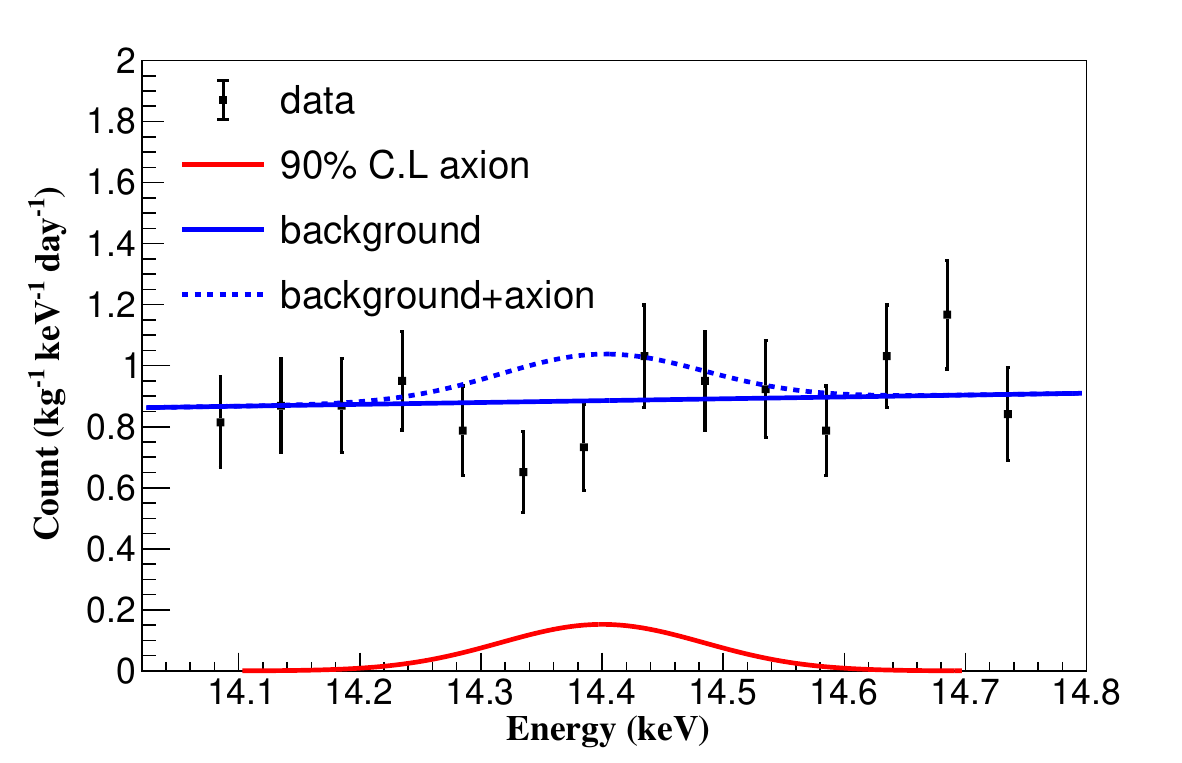}
  \caption{\label{fig:fe_fit} The bulk data (black data points) and the background assumption (solid blue line), as well as the 90\% C.L $^{57}$Fe result (solid red line). The dashed blue line represents the background $+$  90\% C.L signal.  }
 \end{figure}

\begin{figure}[htb]
 \includegraphics[width=1.0\linewidth]{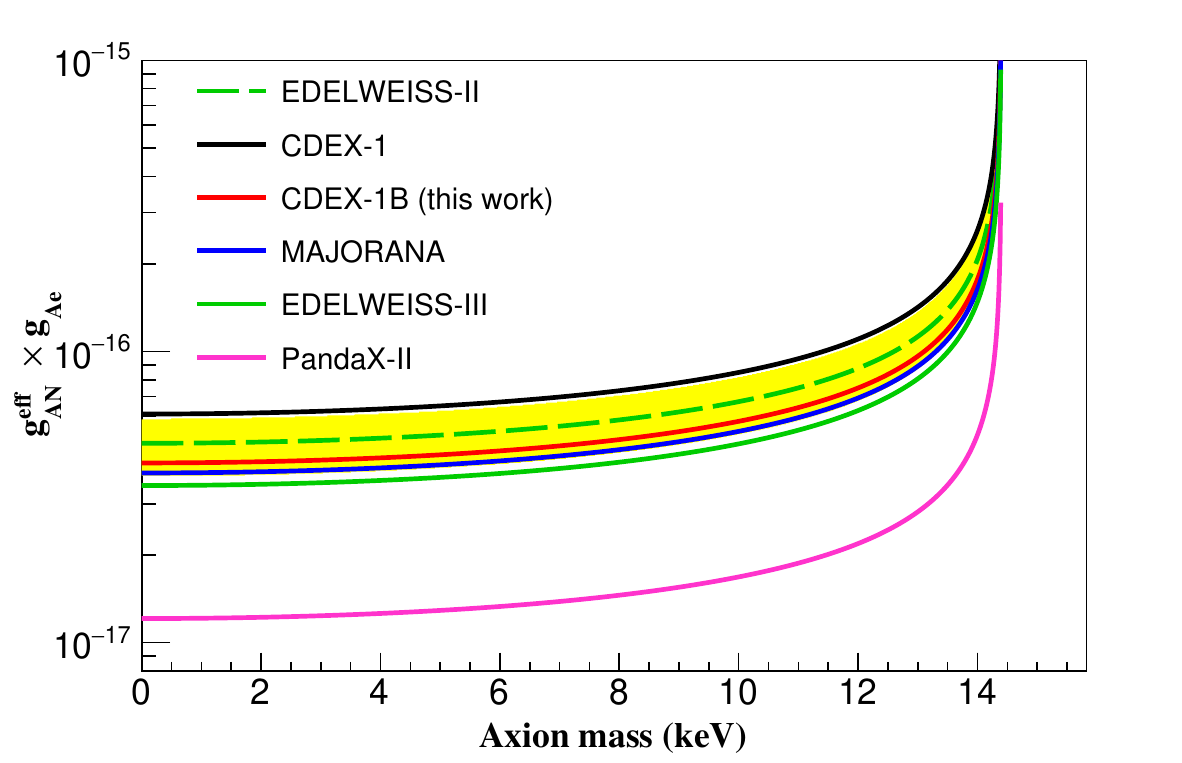}
  \caption{\label{fig:fe57} 90\% C.L. upper limit for the model independent coupling of $g_{AN}^{\rm eff}{\times}g_{Ae}$ of $^{57}$Fe 14.4 keV solar axion (solid red line), compared with CDEX-1A \cite{c1a_axion}, EDELWEISS-II \cite{axion_edelweiss2}, EDELWEISS-III \cite{axion_edelweiss3}, Majorana Demonstrator \cite{axion_majorana} and PandaX-II \cite{axion_pandax2}. The yellow band represents the 1$\sigma$ expected sensitivity.}
 \end{figure}

The signal of solar axions produced in the $^{57}$Fe magnetic transition on the spectrum  is a monochromatic Gaussian peak around 14.4 keV with width determined by resolution, which is about 84 eV ($\sigma$) under this situation. The fitting range is limited to 14.06 keV to 14.76 keV, about $\pm4\sigma$, and a polynomial function is used to described the background in this range. The 90\% C.L result is shown in Fig.~\ref{fig:fe_fit} and the rate of this kind of axion is found to be less than 0.029 counts$\cdot$kg$^{-1}$$\cdot$day$^{-1}$. For a low-mass axion at 0 keV, this result translates to a 90\% C.L. constraint on the coupling:
\begin{equation}
 g_{AN}^{\rm eff}{\times}g_{Ae}<4.14\times10^{-17}.
\end{equation}
Scanning the axion mass from 0 keV to 14.4 keV, we obtained the model-independent limit of $g_{AN}^{\rm eff}\times{g}_{Ae}$ shown in Fig.~\ref{fig:fe57}.
\par Within the framework of a specific axion model, KSVZ or DFSZ, the limits on the couplings $g_{Ae}$ can constrain axion mass $m_{A}$ directly. Using the assumption of parameters mentioned in section II (B), CDEX-1B excludes the mass range 7.3 $\rm eV/c^2$ $< m_A <$ 14.4 $\rm keV/c^2$ for DFSZ axions, and 141.2 $\rm eV/c^2$ $< m_A <$ 14.4 $\rm keV/c^2$ for KSVZ axions.

\subsection{CBRD}
\begin{figure}[htb]
 \includegraphics[width=1.0\linewidth]{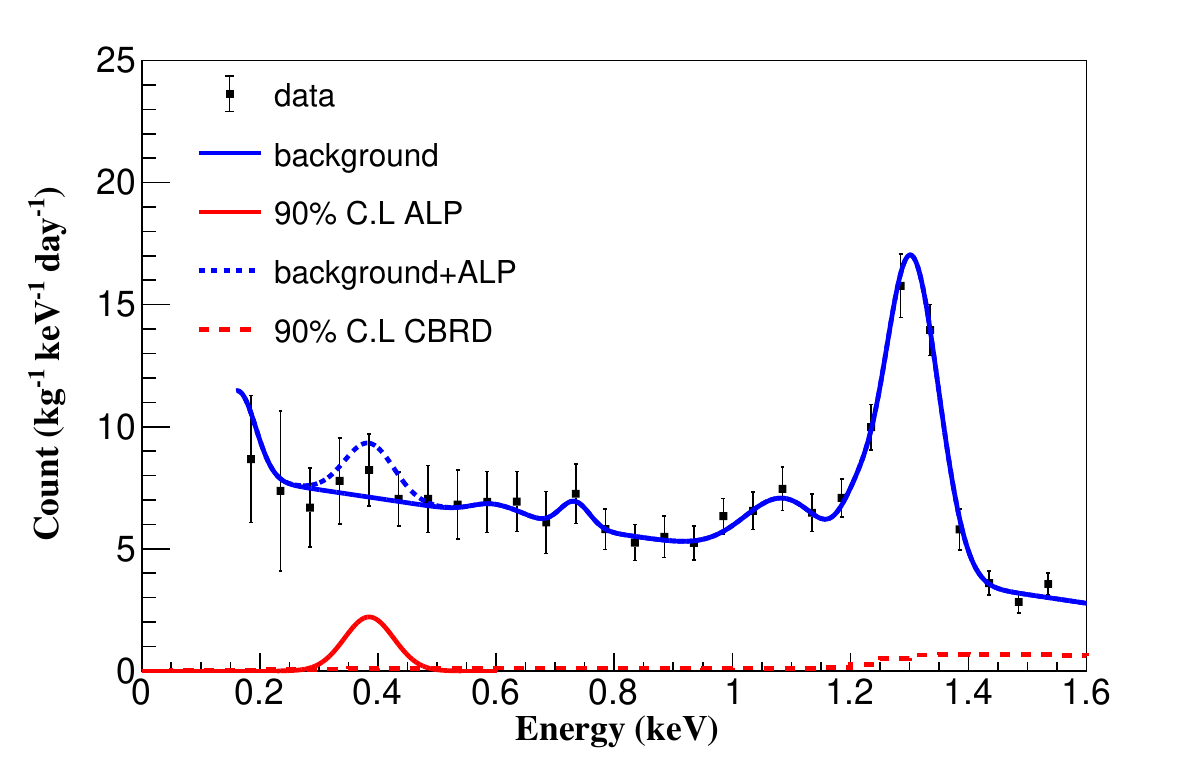}
  \caption{\label{fig:alp_fit} The bulk data (black data points) and the background assumption (solid blue line) below 1.6 keV, as well as the 90\% C.L ALPs result (solid red line) at mass of 385 eV and the 90\% C.L CBRD result (dashed red line). The dashed blue line is the background $+$ 90\% C.L ALP signal. }
 \end{figure}

\begin{figure}[htb]
 \includegraphics[width=1.0\linewidth]{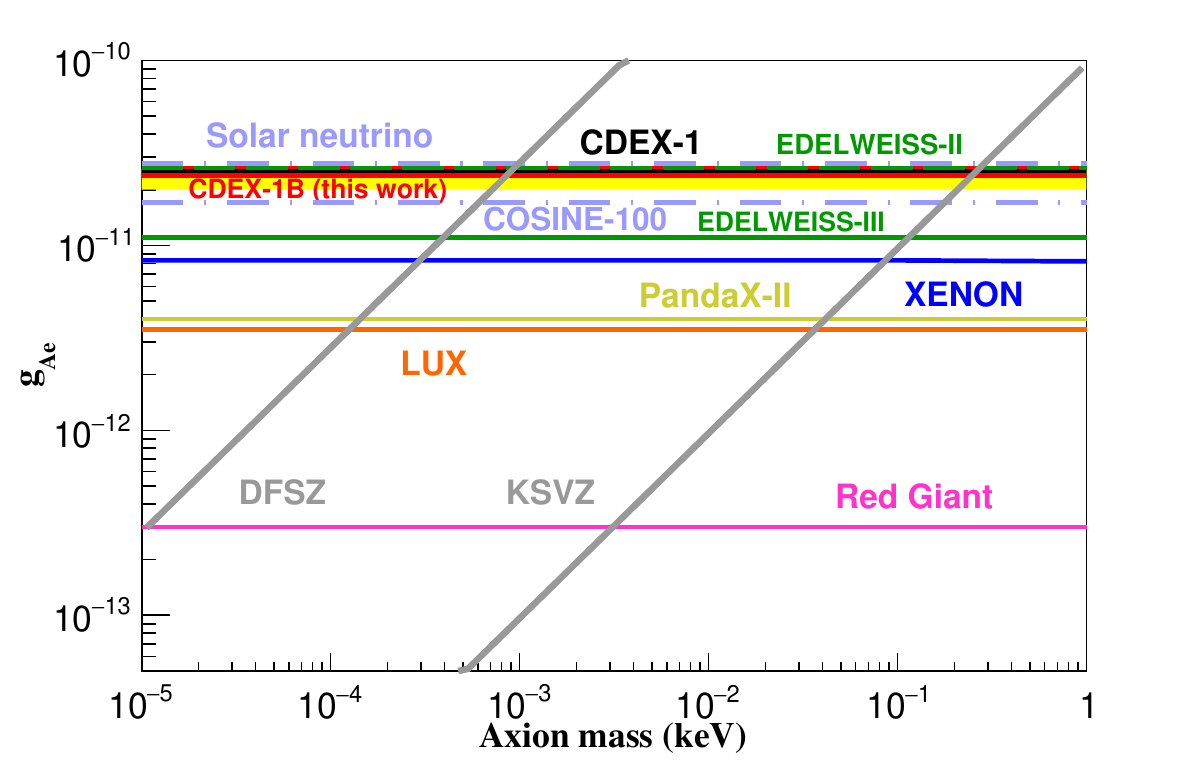}
  \caption{\label{fig:cbrd} The CDEX-1B 90\% C.L. on CBRD solar axions (solid red line), together with astrophysical bounds \cite{cbrd_solar,cbrd_red} and other direct search experiments \cite{axion_edelweiss2,axion_edelweiss3,axion_xenon_2017,axion_pandax2,axion_lux,cos100}. The yellow band represent the 1$\sigma$ expected sensitivity.}
 \end{figure}

For CBRD solar axions, the fitting range is from 0.8 keV to 2.0 keV, and there is a saw-tooth-like profile arising in this energy range which is different from the continuous background. Using the analysis procedure mentioned above, we get the constraints on $g_{Ae}$:
\begin{equation}
  g_{Ae}<2.48\times10^{-11}.
\end{equation}
 Fig.~\ref{fig:alp_fit} depicts the fitting results of 90$\%$ C.L. This result, together with other experimental bounds, is displayed in Fig.~\ref{fig:cbrd}. This result excludes the axion masses $m_A$ $>$ 0.9 eV$\rm /c^2$ in the DFSZ model or $m_A$ $>$ 257.3 eV$\rm /c^2$ in the KFSZ model, which is better than the result of CDEX-1A.

\subsection{Bosonic Dark Matter}
\par
\begin{figure}[htb]
 \includegraphics[width=1.0\linewidth]{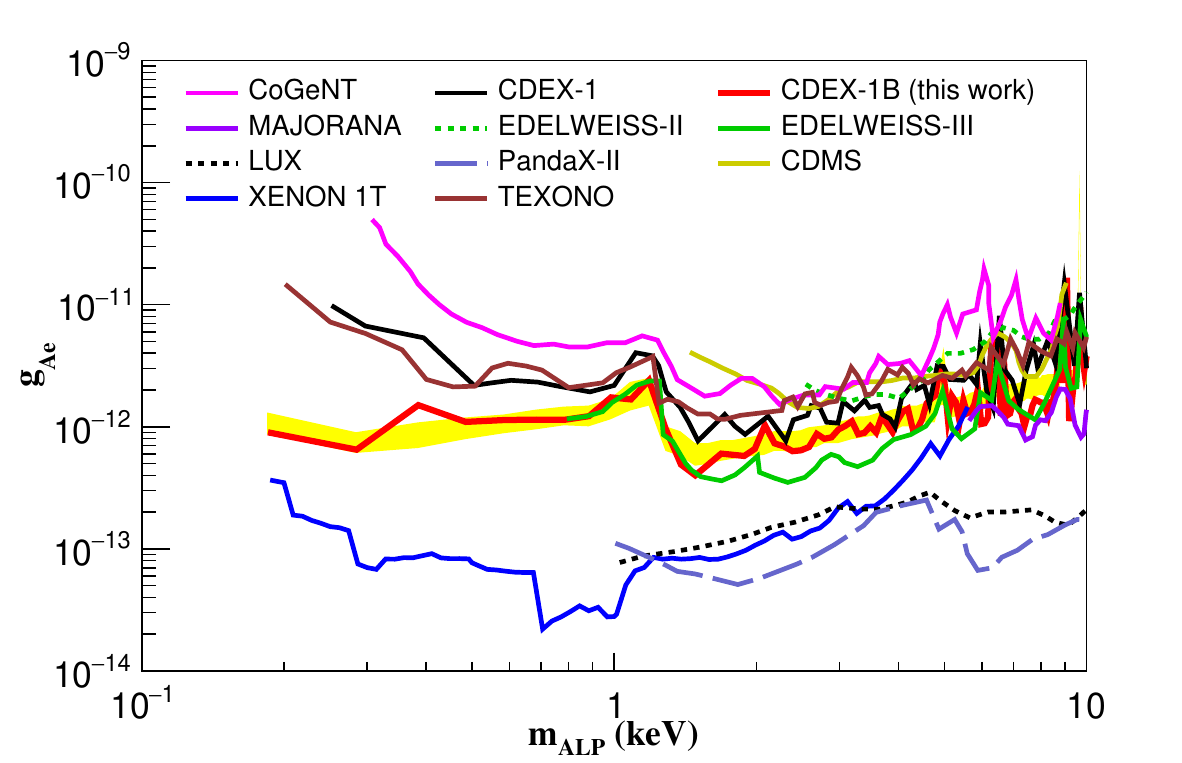}
  \caption{\label{fig:cdex_alp} The CDEX-1B 90\% C.L. upper limit on coupling of ALPs as a function of $m_{\rm ALP}$, together with the constraints set by CDEX-1A \cite{c1a_axion} and other experiments \cite{axion_cogent,axion_cdms,axion_edelweiss2,axion_edelweiss3,axion_pandax2,xenon_1t_axion,axion_lux,axion_majorana,texono}.
  The yellow band represents the 1$\sigma$ expected sensitivity.}
 \end{figure}

\begin{figure}[htb]
 \includegraphics[width=1.0\linewidth]{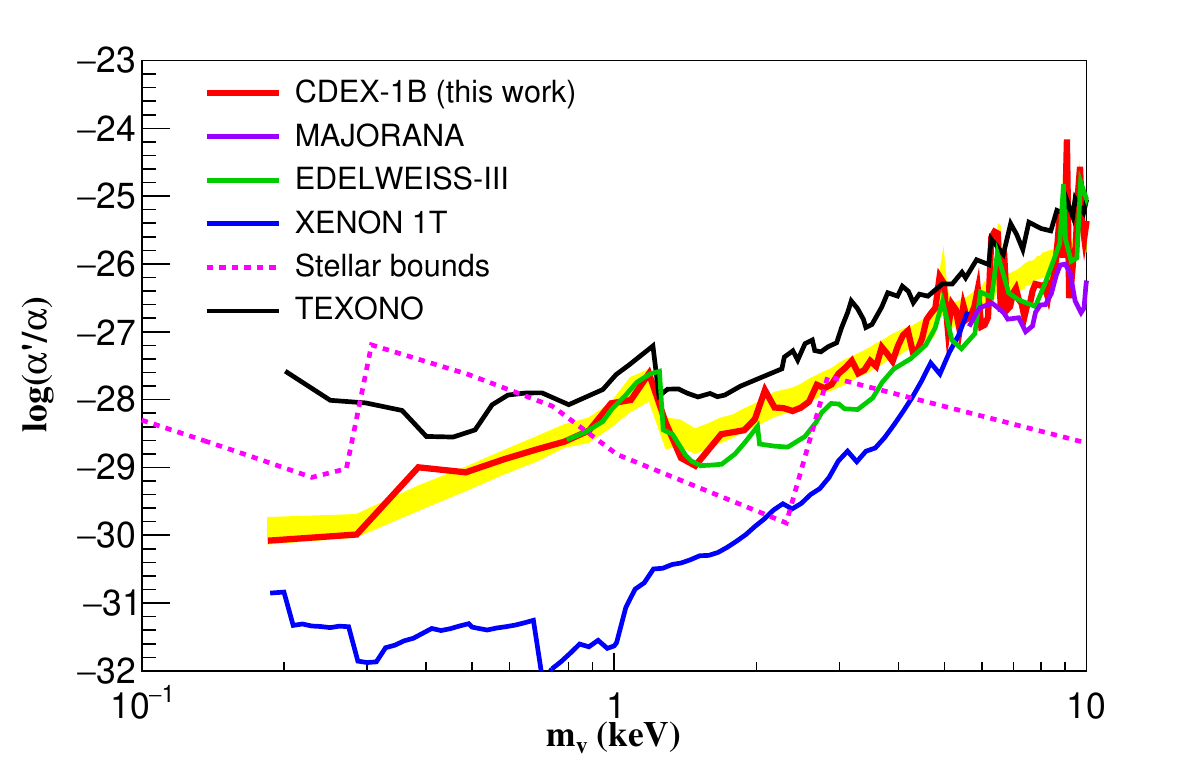}
  \caption{\label{fig:cdex_vec} The 90\% C.L. upper limit on the coupling of vector bosonic DM from CDEX-1B together with the result of EDELWEISS-III \cite{axion_edelweiss3}, Majorana Demonstrator \cite{axion_majorana}, XENON 1T \cite{xenon_1t_axion}, TEXONO \cite{texono} and astrophysical bounds from \cite{darkphoton_anhp}. The yellow band represents the 1$\sigma$ expected sensitivity. }
 \end{figure}

 For bosonic dark matter, the fitting range is from 0.16 keV to 11.66 keV and Fig.~\ref{fig:alp_fit} displays the fitting results at the mass of 385 eV as well as the background model below 1.6 keV. 
Because of the monochromatic signal, better energy resolution and larger exposure, the CDEX-1B gives us much better results of bosonic DM comparing with CDEX-1A. The 90\% C.L. limits on $g_{Ae}$ of ALPs and ${\alpha}'/{\alpha}$ of vector bosonic DM are displayed in Fig.~\ref{fig:cdex_alp} and Fig.~\ref{fig:cdex_vec} respectively. Due to the lower energy threshold, we can extend the first point of exclusion line down to the 185 eV.

\newpage
\section{SUMMARY}
Tighter constraints on the couplings of solar axions and bosonic DM are obtained from CDEX-1B data with an exposure of  737.1 kg-days. Competitive results at the mass of sub-keV of ALPs and vector bosonic DM have been achieved by the help of lower energy threshold and excellent energy resolution measured by the germanium detectors.


These results demonstrate that the profile likelihood ratio method successfully derived the upper limits for our CDEX-1B data in the presence of backgrounds based on the bulk/surface rise-time distribution PDFs. This statistical model takes the main systematic uncertainties, including bulk/surface selection and combined efficiencies, into account through the construction of the likelihood function. The aim of the analysis developed is to provide a reliable statistical forcast of positive signals. 

The CDEX-10 detector array with a target mass of the range 10 kg has provided results on low-mass WIMP searches \cite{cdex102018} and will be installed in a new $~$1700 m$^3$ large LN$_2$ at CJPL-II \cite{cjpl}. In the meantime, the home-made germanium detectors with ultra-low-background electronics are being pursued, which establishes a platform to study the crucial technologies and foreseens to suppress the background.




\begin{acknowledgments}
This work is supported by the National Key Research and Development Program of China (No. 2017YFA0402200), the National Natural Science Foundation of China (No. 11505101, 11725522, 11675088, 11475099 and 11475092), the Fundamental Research Funds for the Central Universities(No. 20822041C4030) and Tsinghua University Initiative Scientific Research Program (Grant No. 20197050007).

The authors of affiliations 5 and 11 participated as members of TEXONO Collaboration.
\end{acknowledgments}



\bibliography{cdex1b_axion}

\end{document}